\documentclass[12pt,a4paper]{article}

\newcommand{\si}{\sigma-\sigma'}
\newcommand{\ta}{\tau-\tau'}
\newcommand{\om}{\omega^{(n)}}

\begin{document}

\begin{flushright}
hep-th/0310044
\end{flushright}
\vspace{1.8cm}

\begin{center}
 \textbf{\Large String Propagators in Time-Dependent and \\
Time-Independent Homogeneous Plane Waves}
\end{center}
\vspace{1.6cm}
\begin{center}
 Shijong Ryang
\end{center}

\begin{center}
\textit{Department of Physics \\ Kyoto Prefectural University of Medicine
\\ Taishogun, Kyoto 603-8334 Japan}  \par
\texttt{ryang@koto.kpu-m.ac.jp}
\end{center}
\vspace{2.8cm}
\begin{abstract}
For a special time-dependent homogeneous plane wave
background that includes a null singularity we construct  
the closed string propagators. We carry out
the summation over the oscillator modes and extract the worldsheet 
spacetime structures of  string propagators specially near the
singularity. We construct the closed string propagators in a 
time-independent smooth homogeneous plane wave background characterized
by the constant dilaton, the constant null NS-NS  field strength and 
the constant magnetic field. By expressing them in terms of the 
hypergeometric function we reveal the background field dependences 
and the worldsheet spacetime structures of string propagators. The 
conformal invariance condition for the constant dilaton plays a role 
to simplify the expressions of string propagators.
\end{abstract} 
\vspace{3cm}
\begin{flushleft}
November, 2003
\end{flushleft}

\newpage
\section{Introduction}

Plane wave spacetimes have led to remarkable developments in our 
understanding of the various aspects of general relativity, string theory
and Yang-Mills theory. The interesting plane wave background with 
maximal supersymmetry \cite{BFHP,BFP} has been produced by the Penrose
limit \cite{RP,RG} on the $AdS_5 \times S^5$ solution of type 
IIB theory. The string theory in this time-independent background is
exactly solvable \cite{RM,MT} so that we
can observe BMN plane-wave/CFT correspondence beyond the supergravity
approximation \cite{BMN}. 

The time-independent plane waves have been
generalized to the time-dependent plane waves in a way which does
not destroy the homogeneity of the metric \cite{PRT,BL}.
There has been an investigation of 
a string theory in a special time-dependent homogeneous
plane wave spacetime with a null singularity, that is
obtained by a Penrose limit of some cosmological, D$p$-brane and 
fundamental string backgrounds. In the light-cone gauge the string 
equation has been solved by using Bessel's function and the quantization
of the string theory has been performed \cite{PRT},
where it is argued that strings pass through the null singular point.
The string theories in the time-dependent plane wave backgrounds 
have been previously studied from various view points 
\cite{HS,RB,VS}.  On the other hand from a direct analysis of 
the Killing equations, the time-dependent homogeneous plane wave metrics
specified by a rotation matrix $f_{ij}$ have been presented and classified
into two families, the regular one generalizing the Cahen-Wallach metrics
with the constant symmetric matrix $A_{ij}$ and the singular one 
generalizing the above special time-dependent plane wave metric 
with a null singularity \cite{BL}. Under some time-dependent coordinate
transformation the former family is changed back 
into the time-independent metric
of regular homogeneous plane wave background specified by 
a time-dependent or time-independent dilaton, the constant rotation
matrix $f_{ij}$, a constant symmetric matrix $k_{ij}$ and a 
constant anti-symmetric matrix $h_{ij}$ associated with the null
NS-NS three-form field strength \cite{BLPT}. 
The closed string theory in this time-independent smooth homgeneous plane
wave background has been solved in the light-cone gauge
and the light-cone Hamiltonian as well
as the string spectrum has been presented. 

There has been also a study of the closed string theory in the 
parallelizable pp-wave background with a constant dilaton
and a constant null NS-NS three-form field strength \cite{SJ}, where it is
shown that the parallelizable pp-wave backgrounds are necessarily
homogeneous plane wave backgrounds, and that a large class of homogenous
plane waves are parallelizable, stating the necessary conditions.
Recently various investigations have been presented with respect to
the homogeneous plane wave backgrounds, the parallelizable pp-wave 
backgrounds and the related background such as the G\"odel 
universe \cite{HT,JF,KY,BML,BHH,FKY,MHS,SY}.

In order to obtain more insights of the string theory in the 
time-dependent homogeneous plane wave background with a null singularity
we will construct closed string propagators in 
this singular background. We will specialize to 
the string propagators with equal worldsheet time and carry
out the string mode summation to examine the behaviors of string
propagators near the singularity, which are compared with the 
flat-like logarithmic behavior. We will study the 
closed string propagation in the time-independent smooth 
homogeneous plane wave background. The string propagator will 
be constructed for the time-independent dilaton
case to be expressed in terms of the hypergeometric function.
The equal-time string propagator produced from it will be 
expanded in the short-distance separations where it is illustrated
how the leading flat-like logarithmic behavior appears together with 
correction terms. The string propagator to the worldsheet time direction
with the worldsheet space-coordinate fixed will be also analyzed.
For these two kinds of propagators the $f_{ij}$ and $h_{ij}$ dependences
will be argued.

\section{String propagators in the time-dependent singular 
homogeneous plane wave background}

We consider the string motion in the $d+2$ dimensional  time-dependent
homogeneous plane wave spacetime with the metric
\begin{equation}
ds^2 = 2dudv - \frac{k}{u^2}(x^i)^2du^2 + (dx^i)^2, 
\hspace{1cm} k = \mbox{const},
\label{tim}\end{equation}
which has an interesting property that there exists a null singularity
at $u = 0$, that is, an initial singularity. This plane wave metric is
produced by a Penrose limit of the FRW metric \cite{BFP}, and of the
near-horizon geometries of D$p$-brane backgrounds $(k=\frac{(7-p)(p-3)}
{16})$ and the fundamental string background $(k=\frac{3}{16})$
\cite{BFP,FIS,SR}. In the light-cone gauge $U = 2\alpha'p^u \tau$
the bosonic part of the closed string action is given by
\begin{equation}
S = - \frac{1}{4\pi \alpha'} \int d\tau \int_0^{\pi}d\sigma
\left(\partial^a X^i \partial_a X^j \delta_{ij} + 
\frac{k}{\tau^2}X_i^2 \right),
\end{equation}
whose equation of motion is solved explicitly in terms of the 
Bessel functions $J_{\nu-\frac{1}{2}}(z), Y_{\nu-\frac{1}{2}}(z)$
as \cite{PRT,FIS}
\begin{eqnarray}
X^i(\sigma,\tau) &=& x_0^i(\tau) + \frac{i}{2} \sqrt{2\alpha'}
\sum_{n=1}^{\infty} \frac{1}{n} \biggl[ Z(2n\tau)( \alpha_n^ie^{2in\sigma}
+  \tilde{\alpha}_n^ie^{-2in\sigma}) \nonumber \\
 &-& Z^*(2n\tau)( \alpha_{-n}^ie^{-2in\sigma}
+  \tilde{\alpha}_{-n}^ie^{2in\sigma}) \biggr]
\label{xep}\end{eqnarray}
with
\begin{eqnarray}
Z(2n\tau) &\equiv& e^{-i\frac{\pi}{2}\nu}\sqrt{\pi n\tau}[ 
J_{\nu-\frac{1}{2}} (2n\tau) - iY_{\nu-\frac{1}{2}}(2n\tau) ], 
\hspace{1cm} \nu \equiv \frac{1}{2}(1 + \sqrt{1-4k}), \nonumber \\
x_0^i(\tau) &=& \frac{1}{\sqrt{2\nu-1}} ( \tilde{x}^i\tau^{1-\nu} +
2\alpha'\tilde{p}^i\tau^{\nu} ),
\end{eqnarray}
where we restrict ourselves to only the region $\tau > 0$ and the case
of $0 < k < \frac{1}{4}$, corresponding to $\frac{1}{2} < \nu < 1$.
Through a relation for the Bessel functions the canonical commutation
relations are shown to be satisfied in the time-independent way although
the commutators for the modes take the standard forms
\begin{eqnarray}
[\alpha_n^i, \alpha_m^j] &=& n \delta^{ij}\delta_{n+m},\hspace{1cm}
[\tilde{\alpha}_n^i, \tilde{\alpha}_m^j] = n \delta^{ij}\delta_{n+m},
\hspace{1cm} [\alpha_n^i, \tilde{\alpha}_m^j] = 0, \nonumber \\
\, [\tilde{x}^i, \tilde{p}^j] &=& i\delta^{ij}.
\label{com}\end{eqnarray}
Taking account of the contribution of $8-d$ spectator dimensions with 
zero-mode momenta $p_s$ we have the light-cone Hamiltonian
\begin{eqnarray}
H &=& - p_u = \frac{p_s^2}{2p_v} + \frac{1}{\alpha'p_v} \mathcal{H},
\nonumber \\
\mathcal{H} &=& \mathcal{H}_0(\tau) + \frac{1}{2}\sum_{n=1}^{\infty}
\left[ \Omega_n(\tau)( \alpha_{-n}^i\alpha_n^i + \tilde{\alpha}_{-n}^i
\tilde{\alpha}_n^i ) - B_n(\tau)\alpha_n^i\tilde{\alpha}_n^i
- B_n^*(\tau)\alpha_{-n}^i\tilde{\alpha}_{-n}^i \right],
\end{eqnarray}
where $\mathcal{H}_0(\tau)$ is the zero-mode part and both 
$\Omega_n(\tau)$ and $B_n(\tau)$ are the involved functions expressed
in terms of the Bessel functions. This Hamiltonian 
includes the non-diagonal terms
specified by $B_n(\tau), B_n^*(\tau)$. The transformation characterizing
a new set of time-dependent string modes $\mathcal{A}_n^i, 
\tilde{\mathcal{A}}_n^i$ as
\begin{eqnarray}
\frac{i}{n}\left( Z(2n\tau)\alpha_n^i -  Z^*(2n\tau)\tilde{\alpha}_{-n}^i
\right) &=& \frac{i}{\sqrt{\omega_n}}\left( e^{-2i\omega_n\tau}
\mathcal{A}_n^i(\tau) -  e^{2i\omega_n\tau}
\tilde{\mathcal{A}}_{-n}^i(\tau)\right), \nonumber \\
\frac{i}{n}\left( \dot{Z}(2n\tau)\alpha_n^i -  \dot{Z}^*(2n\tau)
\tilde{\alpha}_{-n}^i \right) &=& 2\sqrt{\omega_n}\left( 
e^{-2i\omega_n\tau}\mathcal{A}_n^i(\tau) + e^{2i\omega_n\tau}
\tilde{\mathcal{A}}_{-n}^i(\tau)\right)
\end{eqnarray}
with $\omega_n \equiv \omega_n(\tau) = \sqrt{n^2 + k/4\tau^2}$, can
diagonalize the Hamiltonian into
\begin{equation}
\mathcal{H} = \mathcal{H}_0(\tau) + \sum_{n=1}^{\infty} \omega_n(\tau)
[\mathcal{A}_n^{i\dagger}(\tau)\mathcal{A}_n^i(\tau) + 
\tilde{\mathcal{A}}_n^{i\dagger}
(\tau)\tilde{\mathcal{A}}_n^i(\tau)] + h(\tau),
\end{equation}
where $\mathcal{A}_n^{i\dagger}= \mathcal{A}_{-n}^i,
\tilde{\mathcal{A}}_n^{i\dagger}=\tilde{\mathcal{A}}_{-n}^{i\dagger}$
and $h(\tau)$ is a normal ordering c-function \cite{PRT}.
This expression resembles  the Hamiltonian of a free massive
2-d field theory with an effective 
time-dependent mass $\frac{\sqrt{k}}{\tau}$.
Under the transformation the mode expansion of 
$X^i(\sigma,\tau)$ in (\ref{xep}) is changed into
\begin{eqnarray}
X^i(\sigma,\tau) &=& x_0^i(\tau) + \frac{i}{2} \sqrt{2\alpha'}
\sum_{n=1}^{\infty} \frac{1}{\sqrt{\omega_n(\tau)}} \biggl[ 
e^{-2i\omega_n\tau}( \mathcal{A}_n^i(\tau)e^{2in\sigma} +  
\tilde{\mathcal{A}}_n^i(\tau)e^{-2in\sigma}) \nonumber \\
 &-& e^{2i\omega_n\tau}( \mathcal{A}_{-n}^i(\tau)e^{-2in\sigma}
+  \tilde{\mathcal{A}}_{-n}^i(\tau)e^{2in\sigma}) \biggr],
\label{xae}\end{eqnarray}
where $\mathcal{A}_n^i, \tilde{\mathcal{A}}_n^i$ obey the 
following commutation relations in the time-independent way
\begin{equation}
[\mathcal{A}_n^i(\tau), \mathcal{A}_m^{j\dagger}(\tau)] = \delta_{nm}
\delta^{ij}, \hspace{1cm} [\tilde{\mathcal{A}}_n^i(\tau), 
\tilde{\mathcal{A}}_m^{j\dagger}(\tau)] = \delta_{nm}
\delta^{ij}, \hspace{1cm} [\mathcal{A}_n^i(\tau), 
\tilde{\mathcal{A}}_m^{j\dagger}(\tau)] = 0.
\label{aco}\end{equation}

Here we are ready to study the propagator of closed string in the
time-dependent background in the restricted region $0 < u < \infty$.
We construct a closed string propagator
${}_{\alpha}<X^i(\sigma,\tau)X^j(\sigma',\tau')>_{\alpha}$
 where the vacuum is defined
as the Fock space state which is annihilated by $\alpha_n^i,
\tilde{\alpha}_n^i$ with $n > 0$.  Substituting the mode expansion 
(\ref{xep}) and using the commutators for the modes (\ref{com})
we derive a string propagator for the $\alpha$ vacuum
which is a vacuum at $\tau=\infty$
\begin{eqnarray}
{}_\alpha<X^i(\sigma,\tau)X^j(\sigma',\tau')>_{\alpha} &=& - \delta^{ij}
\frac{1}{2\nu-1}\left(\frac{\tau'}{\tau}\right)^{1-\nu}2i\alpha'\tau
\nonumber \\ &+&
\delta^{ij}\alpha'\sum_{n=1}^{\infty}\frac{1}{n}Z(2n\tau)Z^*(2n\tau')
\cos 2n(\si).
\label{pra}\end{eqnarray}
In the $\nu=1$ case corresponding to $k=0$ the zero mode part becomes
$-2i\alpha'\tau$, that is the expression for the flat space.
For large $\tau, Z(2n\tau)$ is expanded as $Z(2n\tau) \cong e^{-2in\tau}
[1 + O(\tau^{-1}]$ so that the string propagator in the large 
$\tau, \;\tau'$ region is approximately given by 
\begin{equation}
{}_\alpha<X^i(\sigma,\tau)X^j(\sigma',\tau')>_{\alpha} \cong
- \frac{\alpha'}{2}\delta^{ij}\left[
\frac{1}{2\nu-1}\left(\frac{\tau'}{\tau}\right)^{1-\nu}\ln z\bar{z}
+ \ln \left(1-\frac{z'}{z}\right) + \ln \left(1-
\frac{\bar{z}'}{\bar{z}}\right) \right],
\label{log}\end{equation}
where we have used a notation as $z=e^{2i(\tau-\sigma)}, \bar{z}=
e^{2i(\tau+\sigma)}, z'=e^{2i(\tau'-\sigma')}$ and $\bar{z}'=
e^{2i(\tau'+\sigma')}$. The non-zero mode part is the same form as
the flat-space theory. Now we analyze the behavior of the string 
propagator in the small $\tau, \tau'$ region.
For simplicity we work at equal worldsheet time and focus on its spatial
dependence. The oscillator part in (\ref{pra}) with $\tau = \tau'$ 
is then expressed as $\delta^{ij}\alpha'\pi f(\sigma,\sigma',\tau)$ with
\begin{equation}
f(\sigma,\sigma',\tau) = \tau \sum_{n=1}^{\infty} 
H^{(2)}_{\nu-\frac{1}{2}}(2n\tau)H^{(1)}_{\nu-\frac{1}{2}}(2n\tau)
\cos 2n(\si),
\label{fth}\end{equation}
where $H^{(1)}_{\nu-\frac{1}{2}}, H^{(2)}_{\nu-\frac{1}{2}}$ are the 
Hankel functions of the first and second kind.
For small $\tau$ the summation over $n$ modes can be approximately 
replaced by the following integral
\begin{equation}
f(\sigma,\sigma',\tau) \cong \int_0^{\infty}dx H^{(2)}_{\nu-\frac{1}{2}}
(2x)H^{(1)}_{\nu-\frac{1}{2}}(2x)\cos \frac{2(\si)x}{\tau}.
\label{int}\end{equation}
Plugging 
\begin{equation}
H^{(2)}_{\nu-\frac{1}{2}}(2x) = \frac{e^{i\nu\pi}J_{\nu-\frac{1}{2}}(2x)
- iJ_{-(\nu-\frac{1}{2})}(2x)}{\cos \nu\pi}, \hspace{1cm}
H^{(1)}_{\nu-\frac{1}{2}}(2x) = H^{(2)*}_{\nu-\frac{1}{2}}(2x)
\end{equation}
into (\ref{int}), we can perform the integral since we are allowed 
to use the following formulas owing to $-\frac{1}{2}<-(\nu -\frac{1}{2})
<0<\nu -\frac{1}{2}$
\begin{eqnarray}
\int_0^{\infty} dx J_{\nu}^2(x) \cos 2ax &=& \left\{ \begin{array}{rl}
\frac{1}{\pi} Q_{\nu-\frac{1}{2}}(1-2a^2), &\mbox{for $0<a<1$}, \\
-\frac{1}{\pi}\sin \nu\pi Q_{\nu-\frac{1}{2}}(2a^2-1),  &\mbox{for $1<a$},
\end{array} \right. \nonumber \\
\int_0^{\infty} dx J_{\nu}(x)J_{-\nu}(x) 
\cos 2ax &=& \left\{ \begin{array}{rl}
\frac{1}{2} P_{\nu-\frac{1}{2}}(2a^2-1), &\mbox{for $0<a<1$}, \\
0, & \mbox{for $1<a$}, \end{array} \right.
\end{eqnarray}
where $P_{\nu-\frac{1}{2}}, Q_{\nu-\frac{1}{2}}$ are the Legendre 
functions of the first and second kind, and the first formula
holds only for $-\frac{1}{2} <\nu$.  For $0 < \frac{|\si|}
{2\tau} < 1$ we have
\begin{equation}
f = \frac{1}{\cos^2\nu\pi}\left[ \frac{1}{2\pi}(Q_{\nu-1}(-y)
+ Q_{-\nu}(-y)) - \frac{\sin \nu\pi}{2} P_{\nu-1}(y) \right]
\end{equation}
with $y = \frac{(\si)^2}{2\tau^2} - 1$, which is further
simplified to be $f = P_{\nu-1}(y)/2\sin \nu\pi$ with
$y<1$, where we have used two relations, $ Q_{\nu}(y) = \pi
(\cos\nu\pi P_{\nu}(y) - P_{\nu}(-y))/2\sin\nu\pi$ 
and $ P_{\nu}(y) = P_{-\nu-1}(y)$.
In the same way for $1 < \frac{|\si|}{2\tau}, \; f$ is 
expressed again in a compact form as 
$f = P_{\nu-1}(y)/2\sin \nu\pi$
but with $y>1$. Thus the oscillator part of string propagator in the small
$\tau = \tau'$ region is represented by a single special function
\begin{equation}
\delta^{ij}\frac{\alpha'\pi}{2\sin\nu\pi} P_{\nu-1}\left(
\frac{(\si)^2}{2\tau^2} - 1 \right).
\end{equation}
Since the Legendre function of the first kind is expressed in terms of
the hypergeometric function, the oscillator part of string propagator
is expressed separately according to the regions of variable as
\begin{eqnarray}
\delta^{ij}\frac{\alpha'\pi}{2\sin\nu\pi} F\left(1-\nu, \nu, 1, 
1-\frac{(\si)^2}{4\tau^2} \right),& 
\mbox{for $\frac{|\si|}{2\tau}<1$}, \label{sma} \\
\delta^{ij}\frac{\alpha'\sqrt{\pi}}{2\sin\nu\pi} (2y)^{\nu-1}
\Biggl[ \frac{\Gamma(\nu-\frac{1}{2})}
{\Gamma(\nu)}F\left(\frac{1-\nu}{2},
\frac{2-\nu}{2}, \frac{3}{2} - \nu, \frac{1}{y^2} \right) & \nonumber \\
+ \frac{\Gamma(\frac{1}{2} -\nu)}{\Gamma(1-\nu)}\frac{1}{(2y)^{2\nu-1}}
F\left(\frac{1+\nu}{2},\frac{\nu}{2}, \frac{1}{2}+\nu, 
\frac{1}{y^2} \right) \Biggr],& \mbox{for $1<\frac{|\si|}{2\tau}$ }.
\label{lar}\end{eqnarray}
From (\ref{sma}) whose hypergeometric function parametrized by $F(\alpha,
\beta, \gamma, z)$ satisfies $\alpha + \beta = \gamma$, we can 
read the short-distance behavior for $\frac{|\si|}{2\tau}\ll 1$
\begin{equation}
\delta^{ij}\alpha' \ln \frac{2\tau}{|\si|},
\label{sho}\end{equation}
while (\ref{lar}) yields the leading large-distance behavior for
$1 \ll \frac{|\si|}{2\tau}$
\begin{equation}
\delta^{ij}\alpha' \frac{\Gamma(\nu-\frac{1}{2})\Gamma(1-\nu)}
{2\sqrt{\pi}}\left(\frac{\tau}{|\si|}\right)^{2(1-\nu)},
\label{pow}\end{equation}
which shows the power damping, owing to the interval 
$\frac{1}{2}<\nu<1$. Thus we have observed that the leading short-distance
logarithmic behavior of the equal-time string propagator in the small
$\tau$ region is similar to the string propagator 
in the flat spacetime. 

Now let us consider a closed string propagator 
${}_{\mathcal{A}}<X^i(\sigma,\tau)X^j(\sigma',\tau)>_{\mathcal{A}}$
with equal worldsheet time
where we have chosen the other vacuum as the Fock space state
which is annihilated by $\mathcal{A}_n^i, \tilde{\mathcal{A}}_n^i$
with $n>0$, and compare it with the equal-time string propagator
for the $\alpha$ vacuum.
 Using the mode expansion (\ref{xae}) together with the 
commutation relations in (\ref{aco}) we obtain an 
equal-time string propagator for the $\mathcal{A}$ vacuum
which is a vacuum at finite $\tau$
\begin{equation}
{}_{\mathcal{A}}<X^i(\sigma,\tau)X^j(\sigma',\tau)>_{\mathcal{A}} =
- \delta^{ij} \frac{2i\alpha'\tau}{2\nu-1} + 
\delta^{ij} \alpha' \sum_{n=1}^{\infty}\frac{1}{\omega_n}\cos 2n(\si),
\end{equation}
which is rewritten by
\begin{equation}
{}_{\mathcal{A}}<X^i(\sigma,\tau)X^j(\sigma',\tau)>_{\mathcal{A}} =
- \delta^{ij}\left(\frac{2i}{\sqrt{1-4k}} + \frac{1}{\sqrt{k}}
\right)\alpha'\tau + \delta^{ij}\frac{\alpha'}{2}
\sum_{n=-\infty}^{\infty} \frac{1}{\omega_n}e^{2in(\si)}.
\label{pam}\end{equation}
Through the Poisson resummation formula it is possible to perform
the mode summation
\begin{eqnarray}
\sum_{n=-\infty}^{\infty} \frac{1}{\omega_n}e^{2in(\si)}
&=& \sum_{l=-\infty}^{\infty}\int_{-\infty}^{\infty}dx 
e^{2\pi ixl}\frac{1}{\sqrt{x^2+m^2}}e^{2ix(\sigma - \sigma')} \nonumber \\
&=& 2K_0(2m|\si|) + 2\sum_{l \neq 0} K_0(2m|l\pi + \si|)
\label{ko}\end{eqnarray}
with $m=\frac{\sqrt{k}}{2\tau}$, where $K_0(x)$ is the modified Bessel
function. The leading short-distance behavior of the string propagator
(\ref{pam}) for $\frac{\sqrt{k}|\si|}{2\tau}\ll 1$ is expressed from the
first term in (\ref{ko}) as
\begin{equation}
\delta^{ij}\alpha' \ln \frac{2\tau}{\sqrt{k}|\si|},
\label{shr}\end{equation}
where we have used a relation between the modified Bessel functions
\begin{equation}
K_0(x) = -I_0(x)\ln \frac{x}{2} + \sum_{k=0}^{\infty}
\frac{\psi(k+1)}{(k!)^2} \left(\frac{x}{2}\right)^{2k}
\label{kol}\end{equation}
with the psi function $\psi (x)$ and $I_0(0)=1$. It is interesting that
this logarithmic behavior shows the same form as (\ref{sho}) that is
given in the $\frac{|\si|}{2\tau} \ll 1$ region. Thus the propagator
(\ref{pam}) with (\ref{ko}), (\ref{kol}) is expanded in
$\sqrt{k}|l\pi + \si|/\tau$ with integer $l$ for the large $\tau$ region.
In the large $\tau$ limit we recover 
the flat-like logarithmic propagator as expected from the 
negligible time-dependent mass. On the other hand we write down the 
large-distance expansion for $\frac{\sqrt{k}|\si|}{2\tau}\gg 1$ that comes
from the first term of (\ref{ko}) as
\begin{equation}
\delta^{ij}\alpha' \sqrt{\frac{\pi\tau}{2\sqrt{k}|\si|}}
e^{-\frac{\sqrt{k}}{\tau}|\si|}
\sum_{n=0}^{\infty}\frac{\Gamma(n+\frac{1}{2})}
{n!\Gamma(-n+\frac{1}{2})}\left(\frac{\tau}{2\sqrt{k}|\si|}\right)^n.
\label{epp}\end{equation}
This exponential damping behavior is compared with the power damping
one observed in (\ref{pow}). For both vacuums the discriminaion
between the short distance and the large distance is specified by the
similar point $|\si| \approx \tau$. Even in the small $\tau$ region
near the singularity both string propagators can show the flat-like
logarithmic behavior in the short space separation restricted by
$|\si| \ll \tau$. The exponential damping in (\ref{epp}) with the
time-dependent mass $\frac{\sqrt{k}}{\tau}$ in the equal-time propagator
resembles the exponential tail of the 
positive energy part of the invariant $\Delta$ function for the massive
scalar field in the space-like separation. In the $\mathcal{A}$
vacuum the mass parameter $m=\frac{\sqrt{k}}{2\tau}$ appears as a
special combination of $\nu$ and $\tau$, while in the $\alpha$ vacuum
the propagator is expressed in terms of the parameter $\nu$.
However, we note that there is a similarity that the string 
propagators in the $\alpha$ vacuum decrease for the large space
separation when $\nu<1$, and those in the $\mathcal{A}$ vacuum also
decrease when $k>0$ that just corresponds to $\nu<1$.

\section{String propagators in the time-independent smooth homogenous
plane wave background}

We turn to the solvable closed string 
theory in the time-independent smooth
homogeneous plane wave backgrounds with a homogeneous NS-NS three-form
field strength and a constant dilaton. We construct a string propagator
in the case of four-dimensional plane wave 
background with metric \cite{BLPT}
\begin{equation}
ds^2 = 2dudv + k_{ij}x^ix^jdu^2 + 2f_{ij}x^idx^jdu + (dx^i)^2,
\end{equation}
with $i=1,2$, where the  constant symmetric parameter $k_{ij}$ is given
by $k_{ij}=k_i\delta_{ij}$ and the constant antisymmetric rotation
parameter $f_{ij}$, that is interpreted as magnetic field,
is taken by $f_{ij} = f\epsilon_{ij}$. 
This metric is supported by a null 2-form NS-NS potential, 
$B_{ui} = h_{ij}x^j$ with $h_{ij}=h\epsilon_{ij}$ and a dilaton.
For the constant dilaton the string sigma model conformal invariance 
condition provides
\begin{equation}
k_1 + k_2 = 2f^2 - 2h^2.
\label{con}\end{equation}
The orthogonal gauge for the worldsheet metric gives the sigma model
 Lagrangian  $L=\frac{1}{4\pi\alpha'}(g+B)_{MN}\partial_+X^M\partial_-X^N$
with $\alpha'=\frac{1}{2}$ where the string embedding coordinates are
given by $X^M = (U,V,X^i)$. The linear but non-diagonal equations for
string transverse coordinates $X^i$ were solved in the light-cone gauge
by using the frequency base ansatz and analyzing a matrix
\begin{equation}
M(\omega,n)= \left(\begin{array}{cc} \omega^2 + k_1-4n^2 & 2if\omega +
4inh \\ -2if\omega - 4inh & \omega^2 + k_2-4n^2 \end{array} \right).
\end{equation}
The solution for the non-zero mode in the expansion of $X^i = 
\sum_{n=-\infty}^{\infty}X_n^i(\tau)e^{2in\sigma}$
with $X_n^i = (X_{-n}^i)^*$ was presented as
\begin{equation}
X_n^i(\tau) = (-1)^i\sum_{J=1}^4 \xi_J^{(n)}m_{1i}(\om_J)e^{i\om_J\tau},
\end{equation}
where the frequencies $\om_J (J=1,\cdots,4)$ are roots of 
\begin{equation}
\omega^4 + (k_1+k_2-4f^2-8n^2)\omega^2 - 16nfh\omega + 
(k_1-4n^2)(k_2-4n^2) - 16n^2h^2 = 0,
\label{fou}\end{equation}
which is given from the constraint $\det M=0$. The matrix $m_{ij}(\om_J)$
is defined as the minor $m_{ij}$ of $M(\omega,n)$ evaluated for 
$\omega=\om_J$. For the zero mode the solution was also derived as
\begin{equation}
X_0^i(\tau) = (-1)^i \sum_{j=1}^{2} [\xi_j^+m_{1i}(\omega_j)
e^{i\omega_j\tau} + \xi_j^-m_{i1}(\omega_j)e^{-i\omega_j\tau}],
\end{equation}
where $(\xi_j^+)^{\dagger} = \xi_j^-$ and 
the pair frequencies $\{\omega_j,-\omega_j\}$ are roots of 
(\ref{fou}) with $n=0$. In order to obey the time-independent canonical
commutation relations the operators $\xi_J^{(n)}, \xi_j^{\pm}$ 
must satisfy
\begin{eqnarray}
[\xi_J^{(-n)}, \xi_J^{(n)}] &=& C_J^{(n)} 
\equiv \frac{1}{m_{11}(\om_J)\prod_{K\neq J}(\om_J - \om_K)},
 \label{can} \\
\,[\xi_j^-, \xi_j^+] &=& C_j \equiv \frac{1}{2m_{11}(\omega_j)\omega_j
\prod_{k\neq j}(\omega_j^2 - \omega_k^2)}
\label{cam}\end{eqnarray}
and the other commutators vanish. When $k_1=k_2=k$, the roots of 
(\ref{fou}) are explicitly obtained by 
\begin{equation}
\{\om_J \} = \{ f \pm \sqrt{f^2 + 4n^2 -k +4 hn}, \; -f \pm 
 \sqrt{f^2 + 4n^2 -k - 4hn} \}.
\label{roo}\end{equation}

Here by substituting the conformal invariance condition 
$k = f^2 - h^2$ (\ref{con}) into (\ref{roo}), we can obtain simple 
expressions without square root and assign the frequencies 
$\om_J$ as
\begin{equation}
\om_1 = |2n + h| + f,\; \om_2 = |2n- h| -f, \; \om_3 = -|2n - h| - f, \;
\om_4 = -|2n + h| + f.
\end{equation}
For the string mode the solution takes the form
\begin{equation}
X^i(\sigma,\tau) = (-1)^i \sum_{n=1}^{\infty}[\sum_{J=1}^{4}\xi_J^{(n)}
m_{1i}(\om_J)e^{i\om_J\tau}e^{2in\sigma} + \sum_{J=1}^{4}\xi_J^{(-n)}
m_{1i}(\omega^{(-n)}_J)e^{i\omega^{(-n)}_J \tau}e^{-2in\sigma}].
\label{non}\end{equation}
For simplicity we assume that $0 < h <2, \; 0 < f$. 
Since there are relations such as $\omega^{(-n)}_1 = - \om_3, 
\omega^{(-n)}_2 = - \om_4, \omega^{(-n)}_3 = - \om_1, \omega^{(-n)}_4
 = - \om_2$ for $n >0$, we label the frequencies $\omega^{(-n)}_J$ in the
last four terms in (\ref{non}) as $\omega^{(-n)}_1 = \omega^{(-n)}_3, 
\omega^{(-n)}_2 = \omega^{(-n)}_4, \omega^{(-n)}_3 = \omega^{(-n)}_1, 
\omega^{(-n)}_4 = \omega^{(-n)}_2$, that is, the labelling of $J=1,2,3,4$
is replaced by that of $J' = 3,4,1,2$. Reshuffling the summation over
$J'$ we rewrite the last four terms in (\ref{non}) as
\begin{equation}
\sum_{J'=1}^{4} \xi_{J'}^{(-n)}m_{1i}(-\om_{J'})
e^{-i\om_{J'}\tau}e^{-2in\sigma},
\label{res}\end{equation}
where $\omega_{J'}^{(-n)}=-\om_{J'}$.  From (\ref{can})
every $C_J^{(n)}$ for $n >0$ is evaluated as
\begin{eqnarray}
C_1^{(n)} &=& \frac{1}{16(f+h)^2(2n+f)^2(2n+h)}, \;
C_2^{(n)} = \frac{1}{16(f+h)^2(2n-f)^2(2n-h)}, \nonumber \\
C_3^{(n)} &=& - \frac{1}{16(f-h)^2(2n+f)^2(2n-h)}, \; 
C_4^{(n)} = - \frac{1}{16(f-h)^2(2n-f)^2(2n+h)},
\label{cn}\end{eqnarray}
where $m_{11}(\omega) = \omega^2 + f^2 -h^2 - 4n^2$ has been used but here
the other components of the minor are written 
down as $m_{12}(\omega) = - m_{21}
(\omega) = -2if\omega - 4inh, m_{22}(\omega) = m_{11}(\omega)$
for convenience. Since $C_1^{(n)} > 0, C_2^{(n)} > 0, 
C_3^{(n)} < 0, C_4^{(n)} < 0$ for 
$n >0 $, the operators such as $\xi_1^{(n)}, \xi_2^{(n)}, 
\xi_3^{(-n)}, \xi_4^{(-n)}$ in the expansion (\ref{non}) with 
(\ref{res}) are associated with the creation operators, while 
$\xi_1^{(-n)}, \xi_2^{(-n)}, \xi_3^{(n)}, \xi_4^{(n)}$ are
regarded as the annihilation operators up to the normalizations.
For the $h > 2$ or the $h < 0$ case, from (\ref{cn}) the assignments of
the creation operators or the annihilation operators are changed but
the analysis remains essentially the same. 
 
Now we construct a closed string propagator $<X^i(\sigma,\tau)
X^j(\sigma',\tau')>$ where the vacuum is defined as the Fock space state
which is annihilated by $\xi_1^{(-n)}, 
\xi_2^{(-n)}, \xi_3^{(n)}, \xi_4^{(n)}$
with $n > 0$. The diagonal (1,1) component of the propagator for the
string non-zero mode is shown to be equal to the (2,2) component as
\begin{eqnarray}
&<X^1(\sigma,\tau)X^1(\sigma',\tau')> = <X^2(\sigma,\tau)
X^2(\sigma',\tau')>& \nonumber \\ 
&= \frac{1}{4}\sum_{n=1}^{\infty} \biggl[ \biggl(\frac{1}{2n+h}
e^{-i(2n+h+f)(\ta)} + \frac{1}{2n-h}e^{-i(2n-h-f)(\ta)} \biggr) 
e^{-2in(\si)}& \nonumber \\ &+ \bigg(\frac{1}{2n-h}e^{-i(2n-h+f)(\ta)}
+ \frac{1}{2n+h}e^{-i(2n+h-f)(\ta)} \biggr)e^{2in(\si)} \biggr],&
\label{phf}\end{eqnarray}
where we have observed that suitable cancellations occur to yield the
simplified coefficients. In calculating the (2,2) component we have 
estimated $m_{12}(-\om_J)$ in the following way. As seen in the steps
from (\ref{non}) to (\ref{res}), $m_{12}(-\om_J)$ is regarded as a
component of the minor $m_{ij}$ for $M(\omega,-n)$ evaluated at
$\omega = -\om_J$. Therefore $m_{12}(-\om_J) = -2if\omega - 
4i(-n)h|_{\omega=-\om_J} = m_{21}(\om_J)$.
The non-diagonal (2,1) component of the propagator is also 
equal to the (1,2) component except for the sign as
\begin{eqnarray}
&<X^2(\sigma,\tau)X^1(\sigma',\tau')> = -<X^1(\sigma,\tau)
X^2(\sigma',\tau')> &\nonumber \\ 
&= -\frac{i}{4}\sum_{n=1}^{\infty} \biggl[ \biggl(\frac{1}{2n+h}
e^{-i(2n+h+f)(\ta)} - \frac{1}{2n-h}e^{-i(2n-h-f)(\ta)} \biggr) 
e^{-2in(\si)}& \nonumber \\ &+ \bigg(\frac{1}{2n-h}e^{-i(2n-h+f)(\ta)}
- \frac{1}{2n+h}e^{-i(2n+h-f)(\ta)} \biggr)e^{2in(\si)} \biggr].&
\label{ndi}\end{eqnarray}
If we do not impose (\ref{con}) but use (\ref{roo}) as the string 
frequencies for general values of $f, h$ and $k$, where the dilaton 
has time-dependence such that the conformal 
invariance condition is satisfied,
then the string propagator has the mode expansion with the
fractional and complicated coefficients, compared with the
simple forms $\frac{1}{2n\pm h}$ in (\ref{phf}), (\ref{ndi}).

There remains a task to analyze the zero-mode contributions.
For the zero mode the equation (\ref{fou}) with $n=0$ gives the roots
$\{ \pm|f - h|, \pm(f + h) \}$. The choice of $\omega_j, j=1,2$ is
taken as $\omega_1= -|f - h|, \omega_2 = f + h$. First we consider the
case $h > f$ which yields $k < 0$ through (\ref{con}), that corresponds
to $0<k<\frac{1}{4}$ for the  time-dependent 
singular homogeneous plane wave metric
(\ref{tim}) owing to the opposite sign of $du^2$.
In this case both $C_j$ are determined from (\ref{cam}) as 
\begin{equation}
C_1 = -\frac{1}{16(f-h)^2f^2h},\hspace{1cm} 
C_2 = \frac{1}{16(f+h)^2f^2h},
\end{equation}
whose signs imply that the operators $\xi_1^+, \xi_2^-$ are
essentially treated as the annihilation operators, while 
$\xi_1^-, \xi_2^+$ as the creation operators.
From this prescription the zero-mode contribution to the string
propagator is obtained in a matrix form as
\begin{equation}
<X_0^i(\tau)X_0^j(\tau')> = \left( \begin{array}{cc}
\cos f(\ta) & \sin f(\ta) \\
-\sin f(\ta) & \cos f(\ta)
\end{array} \right)\frac{e^{-ih(\ta)}}{2h}.
\label{zer}\end{equation}
If we next consider the case $f > h$ implying $k > 0$, then $C_2$ remains
intact but $C_1$ becomes positive as $1/16(f-h)^2f^2h$. 
Since both $C_1$ and $C_2$ are positive, the operators  
$\xi_1^-, \xi_2^-$ are now regarded as the annihilation operators
while $\xi_1^+, \xi_2^+$ as the creation operators. Although this slight
change occurs, the resultant zero-mode string propagator becomes 
the same as (\ref{zer}).

Here we return to the non-zero mode string propagators, (\ref{phf}) and
(\ref{ndi}), which are expressed in terms of the hypergeometric
function as
\begin{eqnarray}
&<X^i(\sigma,\tau)X^i(\sigma',\tau')> =
\frac{1}{4} \biggl[ \frac{1}{2+h}  \biggl(e^{-i(h+f)(\ta)}
\frac{\bar{z}'}{\bar{z}}F\left( \frac{h}{2}+1,1,\frac{h}{2} + 2,
\frac{\bar{z}'}{\bar{z}}\right)& \nonumber \\ 
&+ e^{-i(h-f)(\ta)}\frac{z'}{z}F\left( \frac{h}{2}+1,1,\frac{h}{2} + 2,
\frac{z'}{z}\right) \biggr) + \frac{1}{2-h} \biggl(e^{i(h+f)(\ta)}
\frac{\bar{z}'}{\bar{z}}F\left( -\frac{h}{2}+1,1,-\frac{h}{2} + 2,
\frac{\bar{z}'}{\bar{z}}\right) &\nonumber \\ 
&+ e^{i(h-f)(\ta)}\frac{z'}{z}F\left( -\frac{h}{2}+1,1,
-\frac{h}{2} + 2,\frac{z'}{z}\right) \biggr) \biggr] &
\label{hyp}\end{eqnarray}
and
\begin{eqnarray}
&<X^2(\sigma,\tau)X^1(\sigma',\tau')> =
-\frac{i}{4} \biggl[ \frac{1}{2+h}  \biggl(e^{-i(h+f)(\ta)}
\frac{\bar{z}'}{\bar{z}}F\left( \frac{h}{2}+1,1,\frac{h}{2} + 2,
\frac{\bar{z}'}{\bar{z}}\right)& \nonumber \\
&- e^{-i(h-f)(\ta)}\frac{z'}{z}F\left( \frac{h}{2}+1,1,\frac{h}{2} + 2,
\frac{z'}{z}\right) \biggr) - \frac{1}{2-h} \biggl(e^{i(h+f)(\ta)}
\frac{\bar{z}'}{\bar{z}}F\left( -\frac{h}{2}+1,1,-\frac{h}{2} + 2,
\frac{\bar{z}'}{\bar{z}}\right) &\nonumber \\
&- e^{i(h-f)(\ta)}\frac{z'}{z}F\left( -\frac{h}{2}+1,1,
-\frac{h}{2} + 2,\frac{z'}{z}\right) \biggr) \biggr], &
\label{hyn}\end{eqnarray}
where the same notations as in (\ref{log}) have been used.
We use the expansion formula of the hypergeometric function
$F(\alpha,\beta,\alpha + \beta,z)$ about $z=1$ to express the 
diagonal string propagator (\ref{hyp}) with $\tau=\tau'$ as
the expansion in $\sin \Delta\sigma$ with 
$\Delta\sigma = \si$
\begin{eqnarray}
&<X^i(\sigma,\tau)X^i(\sigma',\tau)> = F(h) + F(-h), &\nonumber \\
&F(h) =  \frac{1}{8}\sum_{n=0}^{\infty} \frac{\Gamma(\frac{h}{2} +1 +n)}
{\Gamma(\frac{h}{2} +1 )\Gamma(n+1)} \biggl[
2\left( \psi(n+1) - \psi(\frac{h}{2} +1 +n) \right) 
\cos\left( (n+2)\Delta\sigma -\frac{\pi}{2}n \right) &\nonumber \\
&- \ln (1-e^{-2i\Delta\sigma})
e^{-i((n+2)\Delta\sigma -\frac{\pi}{2}n)}  
- \ln (1-e^{2i\Delta\sigma})e^{i((n+2)\Delta\sigma -\frac{\pi}{2}n)}
\biggr](2\sin \Delta\sigma)^n ,&
\end{eqnarray}
where there are no $f$-dependences.
The short-distance behavior at $\sigma \approx \sigma'$ is specified by
the leading $n=0$ term that contains the logarithmic function,
which is compared with (\ref{log}), (\ref{sho}) and (\ref{shr}).
The off-diagonal string propagator (\ref{hyn}) with $\tau=\tau'$
is also described by the expansion in  $\sin \Delta\sigma$
\begin{eqnarray}
&<X^2(\sigma,\tau)X^1(\sigma',\tau)> = G(h) - G(-h),& \nonumber \\
&G(h) = - \frac{i}{8}\sum_{n=0}^{\infty} \frac{\Gamma(\frac{h}{2} +1 +n)}
{\Gamma(\frac{h}{2} +1 )\Gamma(n+1)}  \biggl[
-2i\left( \psi(n+1) - \psi(\frac{h}{2} +1 +n) \right)
\sin\left( (n+2)\Delta\sigma -\frac{\pi}{2}n \right) & \nonumber \\
&- \ln (1-e^{-2i\Delta\sigma})
e^{-i((n+2)\Delta\sigma -\frac{\pi}{2}n)}
+ \ln (1-e^{2i\Delta\sigma})e^{i((n+2)\Delta\sigma -\frac{\pi}{2}n)}
 \biggr](2\sin \Delta\sigma)^n, &
\end{eqnarray}
which is manifestly real and has also no $f$-dependences. In the 
short-distance limit the leading $n=0$ term does not include 
the logarithmic function. 

Alternatively, from the expressions (\ref{phf}), (\ref{ndi}) and 
(\ref{zer}), the diagonal equal-time string propagator is given by
\begin{equation}
<X^i(\sigma,\tau)X^i(\sigma',\tau)> = \frac{1}{2h} + \frac{1}{2}
\sum_{n=1}^{\infty} \frac{n}{n^2 - \frac{h^2}{4}} \cos 2n(\si)
\label{eqd}\end{equation}
and the non-diagonal one also takes the form
\begin{equation}
<X^2(\sigma,\tau)X^1(\sigma',\tau)> = \frac{1}{4}
\sum_{n=1}^{\infty}\frac{h}{n^2 - \frac{h^2}{4}}\sin 2n(\si).
\label{eqn}\end{equation}
The expression for the non-zero mode in (\ref{eqd}) is suggestively
compared with the real expressions of the equal-time string
propagators in (\ref{fth}) and (\ref{pam}) with (\ref{ko}).
In the small $h$ expansion the diagonal propagator (\ref{eqd}) is
even function of $\si$ so that it is 
approximately expressed as
\begin{eqnarray}
\frac{1}{2h} - \frac{1}{2} \ln (2\sin |\si|) +\frac{h^2}{8}\biggl[
\zeta(3) + 2(\si)^2\ln 2|\si| \nonumber \\
-3(\si)^2 -\frac{1}{18}(\si)^4 - \frac{1}{1350}(\si)^6 + \cdots \biggr]
+ O(h^4),
\end{eqnarray}
where $\zeta(x)$ is Riemann's zeta function, while the non-diagonal one
(\ref{eqn}) is also expressed as
\begin{equation}
\frac{h}{2}\left[ -(\si)\ln (2|\si|) + (\si) + \frac{1}{18}(\si)^3
+ \cdots \right] + O(h^3),
\end{equation}
which is odd function of $\si$.
In the short-distance limit $\sigma \rightarrow \sigma'$ 
to the space direction the former also shows the logarithmic singular
behavior and the latter vanishes.

In order to focus on the dependence of the string propagator on $f$  we
put $\sigma = \sigma'$ in (\ref{phf}), (\ref{ndi}) to obtain in a 
product form
\begin{eqnarray}
<X^i(\sigma,\tau)X^j(\sigma,\tau')> &=& \left( \begin{array}{cc}
\cos f(\ta) & \sin f(\ta) \\ -\sin f(\ta) & \cos f(\ta)\end{array} 
\right)I(h), \nonumber \\
I(h) &=& \frac{1}{2} \sum_{n=1}^{\infty}\left(\frac{e^{-i(2n+h)(\ta)}}
{2n+h} +  \frac{e^{-i(2n-h)(\ta)}}{2n-h} \right)
\label{fma}\end{eqnarray}
for the non-zero mode, where the $f$-dependent part is separated from
the $h$-dependent one. The zero-mode contribution 
continues to be described by (\ref{zer}). The closed string 
propagator at $\sigma=\sigma'$ is similar 
to the open string propagator on the worldsheet boundary
at $\sigma=0$ or $\sigma=\pi$ where the open string attaches to 
the D-brane. This open string propagator leads to the noncommutativity
on the D-brane worldvolume in the presence of a background NS-NS
three-form field strength \cite{SW}. It is 
interesting that the rotation matrix with the
rotation parameter $f$ appears in the non-zero mode propagator
(\ref{fma}) as well as the zero-mode propagator (\ref{zer}).
 This appearance is also seen in the simple
expression of the string propagator for the non-zero mode in the
$f \neq 0, h =0$ case
\begin{equation}
<X^i(\sigma,\tau)X^j(\sigma',\tau')> = -\frac{1}{4}
\left( \begin{array}{cc}\cos f(\ta) & \sin f(\ta) \\ -\sin f(\ta) &
\cos f(\ta)\end{array} \right)\left[ \ln \left(1 - 
\frac{\bar{z}'}{\bar{z}}\right) +  \ln \left(1 - 
\frac{z'}{z}\right) \right]
\end{equation}
with the same notations as in (\ref{log}).
In (\ref{fma}) the diagonal part has the same mode summation as the
off-diagonal part, which is different from the behaviors of the
equal-time propagators, (\ref{eqd}) and (\ref{eqn}).
The mode summation $I(h)$ is separated into the four kinds of terms as
\begin{eqnarray}
&I(h) = \frac{\cos h(\ta)}{2} \sum_{n=1}^{\infty}\frac{n}{n^2 - 
\frac{h^2}{4}}\cos 2n(\ta) + \frac{\sin h(\ta)}{4} \sum_{n=1}^{\infty}
\frac{h}{n^2 - \frac{h^2}{4}}\sin 2n(\ta)& \nonumber \\
&+ i\left[-\frac{\cos h(\ta)}{2} \sum_{n=1}^{\infty}\frac{n}{n^2 - 
\frac{h^2}{4}}\sin 2n(\ta) + \frac{\sin h(\ta)}{4} \sum_{n=1}^{\infty}
\frac{h}{n^2 - \frac{h^2}{4}}\cos 2n(\ta) \right]. &
\end{eqnarray}
Compared with the real equal-time string propagator, the string 
propagator at $\sigma= \sigma'$ includes an imaginary part.
The real part can be expanded in $h$ in the same way as (\ref{eqd}),
(\ref{eqn}), where $\si$ is replaced by $\ta$. In the imaginary part the
mode summation can be carried out to yield a simle form
\begin{equation}
-i \frac{\pi}{4}\left[ \epsilon(\ta)  - \frac{2}{h\pi} \sin h(\ta) \right]
\end{equation}
for a small region $|\ta| < \pi$. In the short-time limit
$\tau \rightarrow \tau'$,  $I(h)$ exhibits the
logarithmic singular behavior but the off-diagonal propagator
vanishes owing to a multiplying factor $\sin f(\ta)$.
In view of (\ref{eqn}) and (\ref{fma}) with (\ref{zer}) we note that the
non-diagonal propagator with $\tau=\tau'$ is proportional to $h$,
while the non-diagonal propagator with $\sigma=\sigma'$ decreases 
as the magnetic field parameter $f$ becomes much smaller.

\section{Conclusion}

We have constructed the closed string propagators in the special 
time-dependent homogeneous plane wave 
background with a null singularity which
is characterized by a parameter $k$ that is related with $\nu$.
We have carried out the mode summation for the equal-time string 
propagators and presented the simple and analytic expressions in terms of
the the Legendre function of the first kind for the $\alpha$ vacuum,
and the modified Bessel function for the $\mathcal{A}$ vacuum.
For the short separation to the worldsheet space direction, both the
equal-time string propagaotors show the same flat-like logarithmic 
behavior in the leading term. For the large separation the string
propagator in the $\alpha$ vacuum decreases as the power damping only
when $\nu<1$, while the $\mathcal{A}$ vacuum yields the exponential
damping for $k>0$. We have observed an interesting coincidence
between  $\nu<1$ and $k>0$. We have demonstrated that as we  
approach to the singular point more closely, 
the discriminated short-distance region becomes narrower, 
where the string propagators show the leading flat-like
logarithmic behavior. 

In the time-independent smooth homogeneous plane wave backgrounds 
specified by the constant null NS-NS field strength parameter $h$,
the constant magnetic field parameter $f$ and the constant dilaton,
we have constructed the closed string propagator in four dimensions.
We have observed that the conformal invariance condition for the constant
dilaton makes the string propagator tractable and fairly simplified.
The obtained string propagator is described by the hypergeometric function
so that we can make use of an expansion formula of it to express
the equal-time string propagator as the power expansion in
$\sin (\si)$ with the coefficient specified by $h$ only. 
We have presented an alternative expression expanded in $h$ with
the coefficient specified by $\si$. In these two reciprocal expressions
the diagonal propagators include the logarithmic term at the leading
order in each expansion. Analyzing the 
closed string propagator at $\sigma=\sigma'$ we have
demonstrated that it is expressed by a product of two parts, one part
including the $f$ dependence is represented by a single 
rotation matrix and the other part is expanded in $h$ with the
coefficient specified by $\ta$.
We have shown that the presence of the small but finite NS-NS field
strength provides the non-zero contribution to the non-diagonal
equal-time string propagator, while the non-zero contribution to the
non-diagonal string propagator at $\sigma=\sigma'$ is generated
when the magnetic field is turned on.
In deriving the equal-time string propagators 
in the two kinds of homogeneous plane
wave backgrounds we have observed that there appears a critical
point distinguishing the short-distance region from the
large-distance region for the  
time-dependent singular homogeneous plane wave
background, while there is not such a point for the time-independent
smooth homogeneous plane wave background.


\begin{thebibliography}{99}
\bibitem{BFHP} M. Blau, J. Figueroa-O'Farrill, C. Hull and 
G. Papadopoulos,`` A new maximally supersymmetric background 
of IIB superstring theory, " JHEP \textbf{01} (2002) 
047 [hep-th/0110242];`` Penrose limits and maximal supersymmetry, 
" Class. Quant. Grav. \textbf{19} (2002) L87 [hep-th/0201081].
\bibitem{BFP} M. Blau, J. Figueroa-O'Farrill and G. Papadopoulos,
`` Penrose limits, supergravity and brane dynamics, " Class. Quant. Grav.
\textbf{19} (2002) 4753 [hep-th/0202111].
\bibitem{RP} Penrose, ``Any space-time has a plane wave as a limit, "
in Differential Geometry and Relativity, Reidel, Dordrecht, 1976,
pp.271-275.
\bibitem{RG} R. G\"uven, `` Plane wave limits and T-duality, " 
Phys. Lett. \textbf{B482} (2000) 255 [hep-th/0005061].
\bibitem{RM} R.R. Metsaev, `` Type IIB Green-Schwarz superstring 
in plane wave Ramond-Ramond background, " Nucl. Phys. 
\textbf{B625} (2002) 70 [hep-th/0112044].
\bibitem{MT} R.R. Metsaev and A.A. Tseytlin, `` Exact solvable model
of superstring in plane wave Ramond-Ramond background, " Phys. Rev.
\textbf{D65} (2002) 126004 [hep-th/0202109].
\bibitem{BMN} D. Berenstein, J. Maldacena and H. Nastase, `` Strings
in flat space and pp waves from N=4 super Yang Mills, "
JHEP \textbf{04} (2002) 013 [hep-th/0202021].
\bibitem{PRT} G. Papadopoulos, J.G. Russo and A.A. Tseytlin,
`` Solvable model of strings in a time-dependent plane-wave background, "
Class. Quant. Grav. \textbf{20} (2003) 969 [hep-th/0211289].
\bibitem{BL} M. Blau and M. O'Loughlin, `` Homogeneous plane waves, "
Nucl. Phys. \textbf{B654} (2003) 135 [hep-th/0212135].
\bibitem{HS} G.T. Horowitz and A.R. Steif, `` Strings in strong 
gravitational fields, " Phys. Rev. \textbf{D42} (1990) 1950;
`` Space-time singularities in string theory, " 
Phys. Rev. Lett. \textbf{64} (1990) 260.
\bibitem{RB} R. Brooks, `` Plane wave gravitons, curvature singularities
and string physics, " Mod. Phys. Lett. \textbf{A6} (1991) 841.
\bibitem{VS} H.J. de Vega and N. Sanchez, `` Strings falling into
space-time singularities, Phys. Rev. \textbf{D45} (1992) 2783;
H.J. de Vega, M. Ramon Medrano and N. Sanchez, 
`` Classical and quantum strings near space-time singularities:
Gravitational plane waves with arbitrary polarization, " 
Class. Quant. Grav. \textbf{10} (1993) 2007; N. Sanchez, 
`` Classical and quantum strings in plane waves, shock waves and
space-time singularities: synthesis and new results, " hep-th/0302214.
\bibitem{BLPT} M. Blau, M. O'Loughlin, G. Papadopoulos and A.A. Tseytlin,
`` Solvable models of strings in homogenous plane wave backgrounds, "
hep-th/0304198.
\bibitem{SJ} D. Sadri and M.M. Seikh-Jabbari, `` String theory on
parallelizable pp-waves, " JHEP \textbf{06} (2003) 005 [hep-th/0304169].
\bibitem{HT} T. Harmark and T. Takayanagi, `` Supersymmetric G\"odel
universes in string theory, " Nucl. Phys. 
\textbf{B662} (2003) 3 [hep-th/0301206].
\bibitem{JF} J. Figueroa-O'Farrill, `` On parallelisable 
NS-NS backgrounds, " Class. Quant. Grav. \textbf{20} (2003) 
3327 [hep-th/0305079].
\bibitem{KY} T. Kawano and S. Yamaguchi, `` Dilatonic parallelizable
NS-NS backgrounds, " Phys. Lett. \textbf{B568} (2003) 78 [hep-th/0306038].
\bibitem{BML} M. Blau, P. Meessen and M. O'Loughlin, `` G\"odel,
Penrose, anti-Mach: Extra supersymmetries of time-dependent plane waves,
JHEP \textbf{09} (2003) 072 [hep-th/0306161].
\bibitem{BHH} D. Brace, C.A. Herdeiro and S. Hirano,`` Classical and
quantum strings in compactified pp-waves and G\"odel type universes, "
hep-th/0307265.
\bibitem{FKY} J. Figueroa-O'Farrill, T. Kawano and S. Yamaguchi,
`` Parallelisable heterotic backgrounds, " hep-th/0308141.
\bibitem{MHS} M. Hssaini and M.B. Sedra, `` Type IIB string 
backgrounds on parallelizable pp-waves and conformal Liouville
theory, " hep-th/0308187.
\bibitem{SY} M. Sakaguchi and K. Yoshida, `` M-theory on a 
time-dependent plane-wave, " hep-th/0309025.
\bibitem{FIS} H. Fuji, K. Ito and Y. Sekino, `` Penrose limit
and string theories on various brane backgrounds, " JHEP \textbf{11}
(2002) 005 [hep-th/0209004].
\bibitem{SR} S. Ryang, `` Penrose limits of branes and marginal
intersecting branes, " Phys. Lett. \textbf{B555} (2003) 
107 [hep-th/0209218].
\bibitem{SW} N. Seiberg and E. Witten, `` String theory and
noncommutative geometry, " JHEP \textbf{09} (1999) 032 [hep-th/9908142].

\end{thebibliography}
\end{document}